\documentclass[aps,pra,groupedaddress,twocolumn, amsmath]{revtex4-2}

\usepackage{graphicx}
\usepackage{dcolumn}
\usepackage{bm}
\usepackage{graphicx}
\usepackage{ulem}
\usepackage[caption=false]{subfig}
\usepackage{xcolor} 
\usepackage{comment}
\usepackage{hyperref}
\usepackage{xfrac}
\usepackage{upgreek}

\begin{document}

\title{
Laser stabilized to a room temperature cavity with AlGaAs coatings reaching $4.2 \times 10^{-17}$ fractional frequency instability}
\author{Chun Yu Ma, Jialiang Yu, Steffen Sauer, Thomas Legero, Daniele Nicolodi, Mona Kempkes, Sofia Herbers, Fritz Riehle and Uwe Sterr}

\affiliation{Physikalisch-Technische Bundesanstalt, Bundesallee 100, 38116 Braunschweig}

\date{\today}

\begin{abstract}
We present a laser system referenced to a room-temperature ultrastable cavity employing crystalline AlGaAs coatings. We demonstrate a fractional frequency instability of 4.2~$\times 10^{-17}$, which is one of the lowest for room temperature systems and surpasses the limit imposed by Brownian noise if  dielectric coatings were employed.
For the first time in a room temperature system we identified the spontaneous fluctuations of the coating birefringence as a leading contribution to frequency instability.
At optimized conditions we achieve an ultrastable cavity with an eigenfrequency that is highly immune to power fluctuations.
As acceleration noise is the main noise contribution, we demonstrated that a feed-forward method can reduce the influence of accelerations on the cavity-stabilized laser frequency by a factor of four.
\end{abstract}


\maketitle

\section{Introduction}
Lasers with ultrahigh frequency stability and long coherence time are indispensable to numerous applications in basic science and in applied technology. They are used as local oscillators for optical atomic clocks \cite{lud15,for26}
or in laboratory investigations of fundamental physics \cite{wie16, ken20}. 
They are also required for obtaining super-stable microwave frequencies \cite{xie17} for a variety of civilian and military applications from radar to improved long range communication \cite{att24}.

Despite of significant efforts towards other technologies \cite{mei09,ols19,lin24} the best frequency stability is achieved by referencing the laser frequency to ultrastable Fabry-Perot cavities. In these systems the frequency instability is fundamentally limited by Brownian noise of high reflectivity dielectric coatings \cite{num04, har06b}. Lower levels of Brownian noise can be obtained with AlGaAs crystalline coatings \cite{col13, yu23a, lee26}. 
Using these coatings, a fractional frequency instability of $2.5 \times 10^{-17}$ has been achieved with a 6-cm long single-crystal silicon cavity at 17\,K \cite{lee26}. 
In comparison to cryogenic systems, optical cavities operated at room temperature are simpler, can be more easily designed to be transportable, and do not require regular maintenance. 
State-of-the-art room-temperature cavities 
reach instability in the mid $10^{-17}$ \cite{hae15a,sch22a,ma24a,par26}.
These systems are limited by the thermal noise of the dielectric coatings. So far the frequency stability enabled by crystalline coatings has not been realized in a room temperature system. 

In this paper we present a room temperature ultrastable cavity with crystalline AlGaAs coatings achieving $4.2 \times 10^{-17}$ fractional frequency instability. 
We present a detailed noise budget that well describes the achieved stability. 
Special to AlGaAs coatings, spontaneous fluctuations of the coatings birefringence significantly contribute to the noise budget, as observed at cryogenic temperatures \cite{yu23a, ked23}. Here, for the first time, we 
characterize this noise source in a room temperature system.

Acceleration noise limits the performance at short averaging time. We demonstrate that, combining active vibration isolation (AVI) and frequency feed forward correction of the laser frequency based on the acceleration measured by a seismometer and a tilt meter, the frequency stability can be improved by up to a factor of ten. 

\section{Setup and Methods}
The system is realized accordingly to a previously reported design \cite{hae15a}. The cavity comprises of two mirrors separated by a 48\,cm long ultra low expansion glass (ULE) spacer. 
The mirrors consist of fused silica substrates (one plane and one concave with 1.0\,m radius of curvature) with bonded crystalline GaAs/Al$_{0.92}$Ga$_{0.08}$As coatings. 
These coatings are birefringent \cite{col13,win21}. The mirrors are oriented to align the birefringent axes and maximize the birefringent splitting, resulting in two polarization eigenmodes separated by 104\,kHz.
The finesse of the resonator at 1542\,nm is $1.290 (2) \times 10^{5}$ and $1.282 (2) \times 10^{5}$ for the $\mathrm{TEM_{00}}$ mode along the slow and fast axis, respectively. ULE rings are attached to the back of the mirrors to compensate for the coefficient of thermal expansion (CTE) mismatch between ULE and fused silica \cite{leg10}. The cavity is surrounded by three heat shields inside the vacuum chamber. It is thermally stabilized to the CTE zero crossing at 297\,K. 
The vacuum chamber with attached input coupling optics is supported by an AVI system \footnote{The Table Stable Ltd. AVI 200-M}.
A three-axis seismometer and a two-axis tilt meter are placed next to the chamber to monitor accelerations acting on the cavity. 
The whole setup is inside a thermo-acoustic enclosure.

The frequency of an erbium doped fiber laser at 1542\,nm is locked to a $\mathrm{TEM_{00}}$ mode of the cavity via Pound-Drever-Hall (PDH) technique \cite{bla01}. We evaluate the frequency stability of this cavity via three-cornered-hat method (TCH) by measuring against two independent lasers at 1542\,nm locked to cryogenic cavities \cite{mat17a, yu23a}. 
Shrinking of the ULE cavity leads to a frequency drift of about 20\,mHz/s.  
Thus the linear drift is removed before TCH analysis. 

\section{Frequency stability}

\begin{figure}
    \includegraphics[width=\columnwidth]{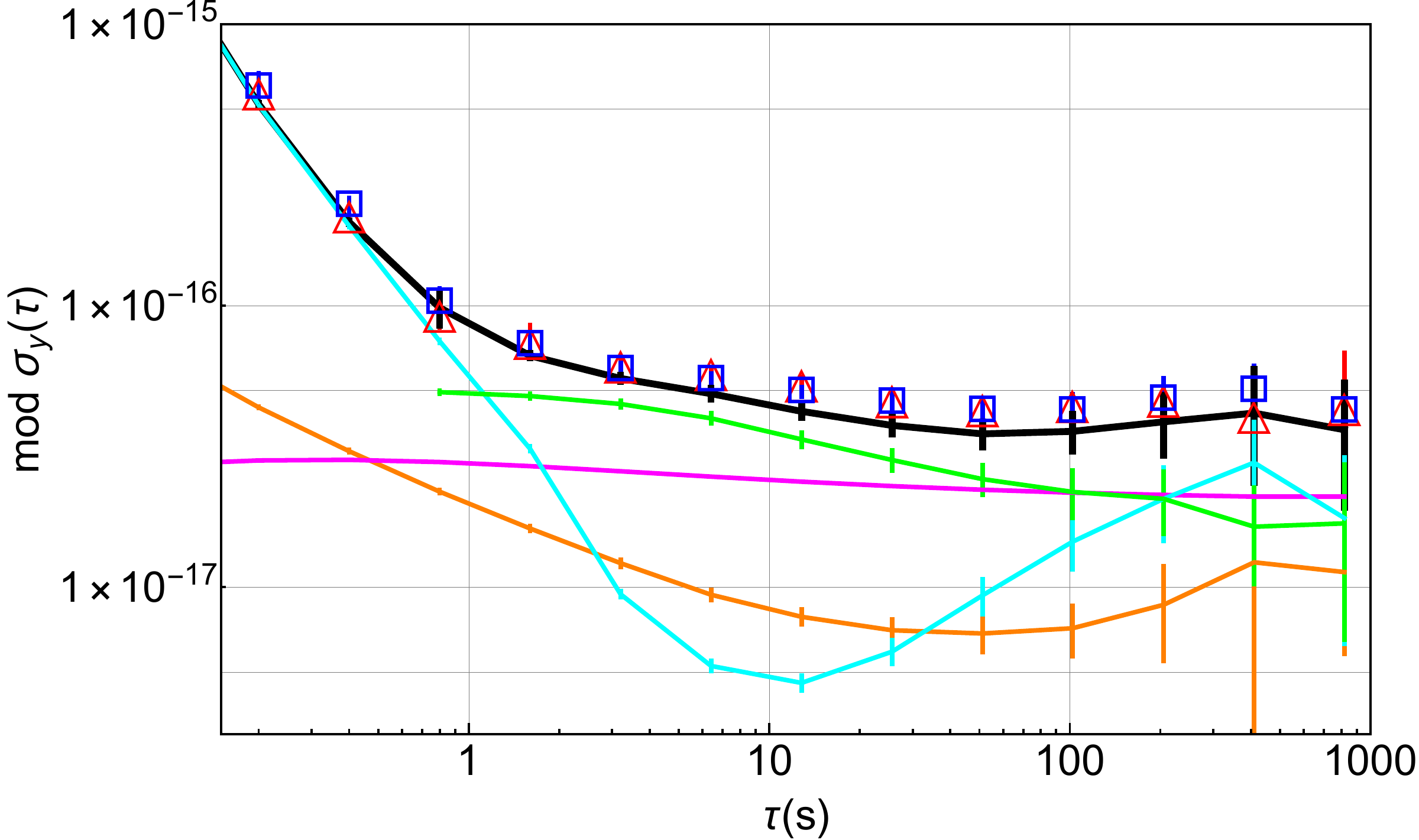}
    \caption{\label{Fig:stability}
    Fractional frequency instability of the cavity stabilized laser evaluated with TCH. The statistical mean of the TCH results is shown for transmission power of $52~\upmu$W (red triangles) and for transmission power of $9~\upmu$W with external illumination (blue boxes). The standard deviation is generally smaller than the markers except for the last data points. The sum of all identified noise sources is shown in black with the error bars showing the statistical uncertainty. The individual components include thermal noise (magenta), acceleration (cyan), noise due to electronics and RAM (orange) and birefringent noise (green).}   
    \end{figure}

The birefringence of AlGaAs coatings can be modified by intracavity light with photon energies below the bandgap of the coating material \cite{yu23a, zhu24, kra25a, ma24a} and with above bandgap illumination (photo-birefringent effects) \cite{wu25c,ma26}. 
Increasing intracavity power or LED intensity decreases the eigenfrequency along the fast axis in this cavity in a nonlinear way.
Absorption of the intracavity light heats up the mirrors and increases the resonance frequencies of the cavity due to photo-thermo-optic effect with a linear sensitivity of 4.5~Hz/$\upmu$W. 
This allows cancellation of the two effects on the laser frequency to small fluctuations of intracavity power at a certain combination of intracavity power and LED illumination \cite{zhu24, kra25a, ma26}.
Since the two temporal responses are different, a perfect cancellation is not possible for all times after a step in laser power.
Using only intracavity power to optimize the response for times up to 10 seconds, the residual sensitivity of our resonator can be kept below 0.5~Hz/$\upmu$W at an intracavity power of 3.5\,W, corresponding to a transmission power of $P_\mathrm{trans} = 52~\upmu$W. 
Under this operating condition, between averaging time of about 10~s and 1000~s a nearly constant instability is achieved (Fig. \ref{Fig:stability}) with the lowest instability of $4.2 (9) \times 10^{-17}$ near 400~s averaging time. 
From the measured power stability and this optimized sensitivity we estimate the contribution of laser power fluctuations to frequency instability to be below $5 \times 10^{-18}$ within the investigated averaging times.
We can achieve a similar low sensitivity to laser power changes at lower intracavity power levels by illuminating the mirrors with a LED (868~nm to 912~nm) installed in front of the vacuum window near the plane mirror \cite{ma26}. 
The measured frequency instability of the stabilized laser with LED illumination shows virtually the same results (Fig. \ref{Fig:stability}). In particular at five times lower intracavity power with LED illumination, the same low instability of $4.2 (6) \times 10^{-17}$ can be achieved between 50 and 100~s. 
The results obtained from the two very different measurement conditions go below the thermal noise floor set by an identical system with dielectric coatings \cite{hae15a}, hence realizing the improvement by using AlGaAs coatings with low Brownian thermal noise at room temperature.

In addition to Brownian thermal noise from the coatings, there is also fundamental Brownian thermal noise from the spacer and mirror substrates, thermo-elastic noise of the mirror substrates and thermo-optic noise from mirror coatings \cite{eva08,cha16,cer01}.
In our system the biggest contributions to thermal noise are the thermo-elastic noise from the fused silica mirror substrates and the Brownian noise from the ULE spacer in the investigated frequency range (Fig. \ref{Fig:stability} magenta).

Both residual amplitude modulation (RAM) \cite{zha14} and electronic noise from locking servos lead to frequency offsets and noise. To suppress RAM, the temperature of the electro-optic modulator (EOM) for phase modulation in PDH locking is stabilized to within a few mK, and remaining RAM is actively canceled by applying a bias voltage to the EOM. 
For largest cancellation efficiency, light for RAM detection is sampled as close as possible to the cavity. 
The optical setup is enclosed inside heat-insulating foams and various optical isolators are employed to suppress parasitic etalons \cite{yu23}. 
To access and to minimize frequency perturbations from RAM, electronics and parasitic etalons, we stabilize the frequency of a second laser to an adjacent fundamental mode of the cavity along the same polarization and compare it with the first laser. 
Assuming uncorrelated noise and similar locking performance of each laser, 1/$\sqrt{2}$ of the beat frequency stability is shown as single laser instability (Fig.~\ref{Fig:stability} orange).

The coating birefringent noise is studied in detail in section IV. We find that the stability between averaging times of 1~s and 100~s is mainly limited by this noise source, similar to cryogenic Si cavities \cite{yu23a, ked23}.

At shorter and longer averaging time, the stability is limited by forces due to accelerations acting on the cavity. 
The accelerations can occur from linear vibrations and from small tilt motions that project a varying part of the gravity along and perpendicular to the spacer length. 
Similarly, the installed seismometer and tilt meter measure acceleration along these axes. 
Because of the different frequency response of the sensors, for averaging times below 10~s for the horizontal directions we use the data from the seismometer and for longer times data from the tilt meter to calculate the acceleration induced noise shown in Fig.~\ref{Fig:stability}. 

The sum of all these noise contributions (Fig.~\ref{Fig:stability} black) is in good agreement with the achieved stability within the error bars. 
This shows we have included the most relevant noise contributions. 
If the slightly higher measured instability compared to the total estimated noise would be due to unknown coating noise, this contribution would be around $2.5 \times 10^{-17}$.
In our system this is below the overall thermal noise of the cavity, different from system at 124~K where an excess noise higher than the thermal noise was observed \cite{yu23a}.
The excess noise scaled to our cavity length would amount to $1.3 \times 10^{-17}$.  

\section{Coating birefringent noise at 297~K}
\begin{figure}
	\includegraphics[width=\columnwidth]{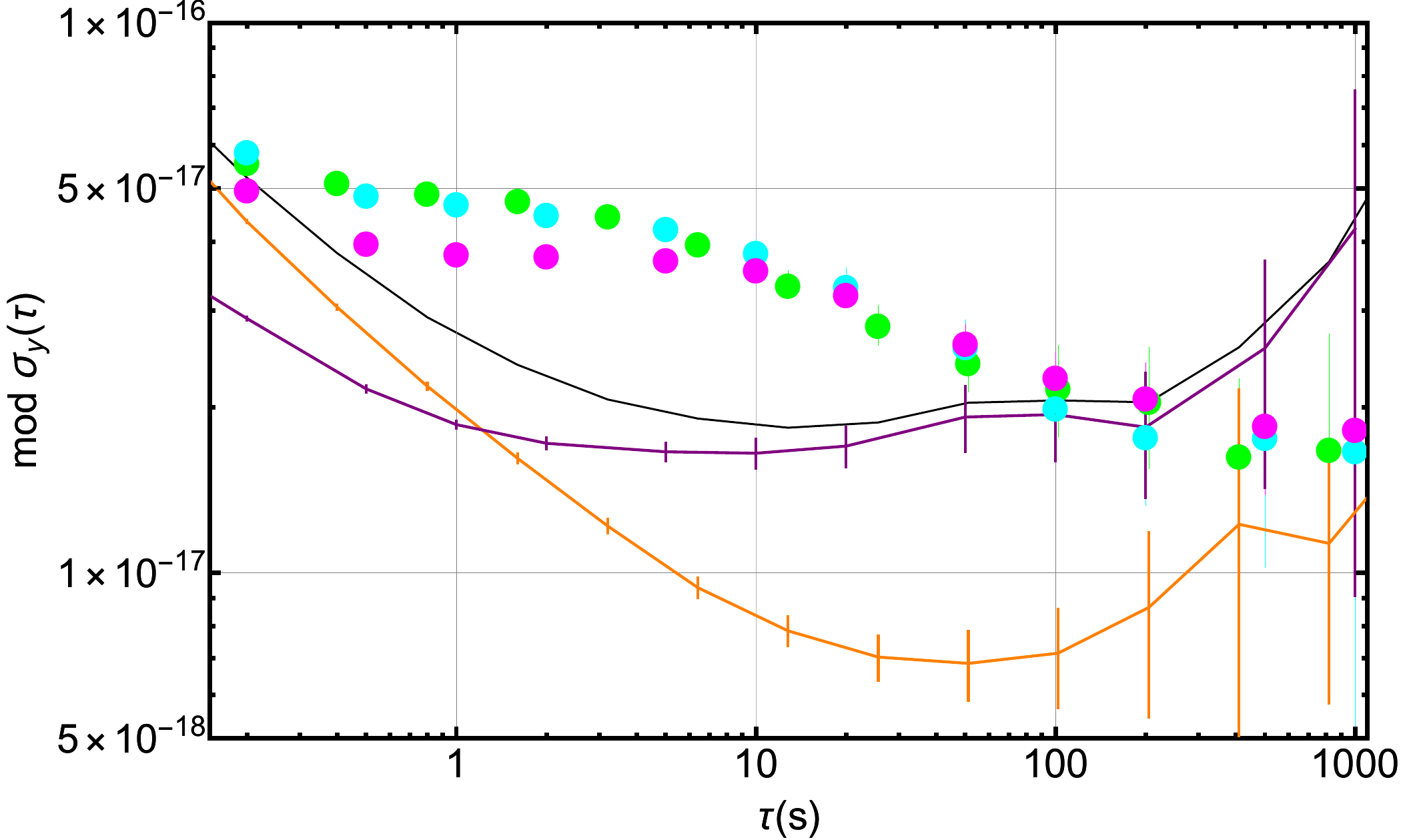}
	\caption{	\label{Fig:birefnoise}
		Contributions of the coating birefringence to the frequency instability of a cavity polarization eigenmode, measured from the beat frequency between two lasers stabilized to orthogonally polarized eigenmodes. The noise is measured under three different transmitted power levels along both polarizations at 52~$\upmu$W (green dots), 27~$\upmu$W (cyan dots) and 13~$\upmu$W (magenta dots) respectively. The lines show the estimated noise from electronics and RAM (orange), the estimated photo-birefringent noise (purple) and their sum (black).    
	}
\end{figure}

Significant spontaneous fluctuations of the coating birefringence have been observed at cryogenic temperatures  \cite{yu23a, ked23}.
To test if there is also coating birefringent noise at room temperature we lock a second laser to an orthogonal polarization and characterize the birefringent line splitting via the beat frequency. 
In this case, on top of technical noises due to RAM and electronics, noises that are anti-correlated including photo-birefringent noise will be seen in this beat frequency. 
The birefringent line splitting is highly sensitive to laser power fluctuations with sensitivities of 25~Hz/$\upmu$W at $P_\mathrm{trans} = 13~\upmu$W to 10~Hz/$\upmu$W at $P_\mathrm{trans} = 52~\upmu$W \cite{ma24a,ma26}.
With these sensitivities and the measured stability of the transmitted power, the contribution of the photo-birefringent noise on a single polarization is calculated using half of the splitting (Fig. \ref{Fig:birefnoise}). 
Contributions due to RAM and electronics were estimated from measurements with parallel polarization as in the previous section. 

We observe birefringent noise between averaging times of 0.5~s and 50~s well above the technical noise contributions. 
In this system we observe a reduction of the noise of 20\% at 1~s and no effect at longer times when $P_\mathrm{trans}$ is reduced by a factor of four.
This is in contrast to the observed birefringent noise at cryogenic temperatures that scales with ${1}/{\sqrt{P_\mathrm{trans}}}$, which would imply a reduction of mod~$\sigma_y$ by 30\% \cite{ked23,yu23a}.

\section{Suppression of tilt motion and vibration effects with feed forward}
\begin{figure}
	\includegraphics[width=\columnwidth]{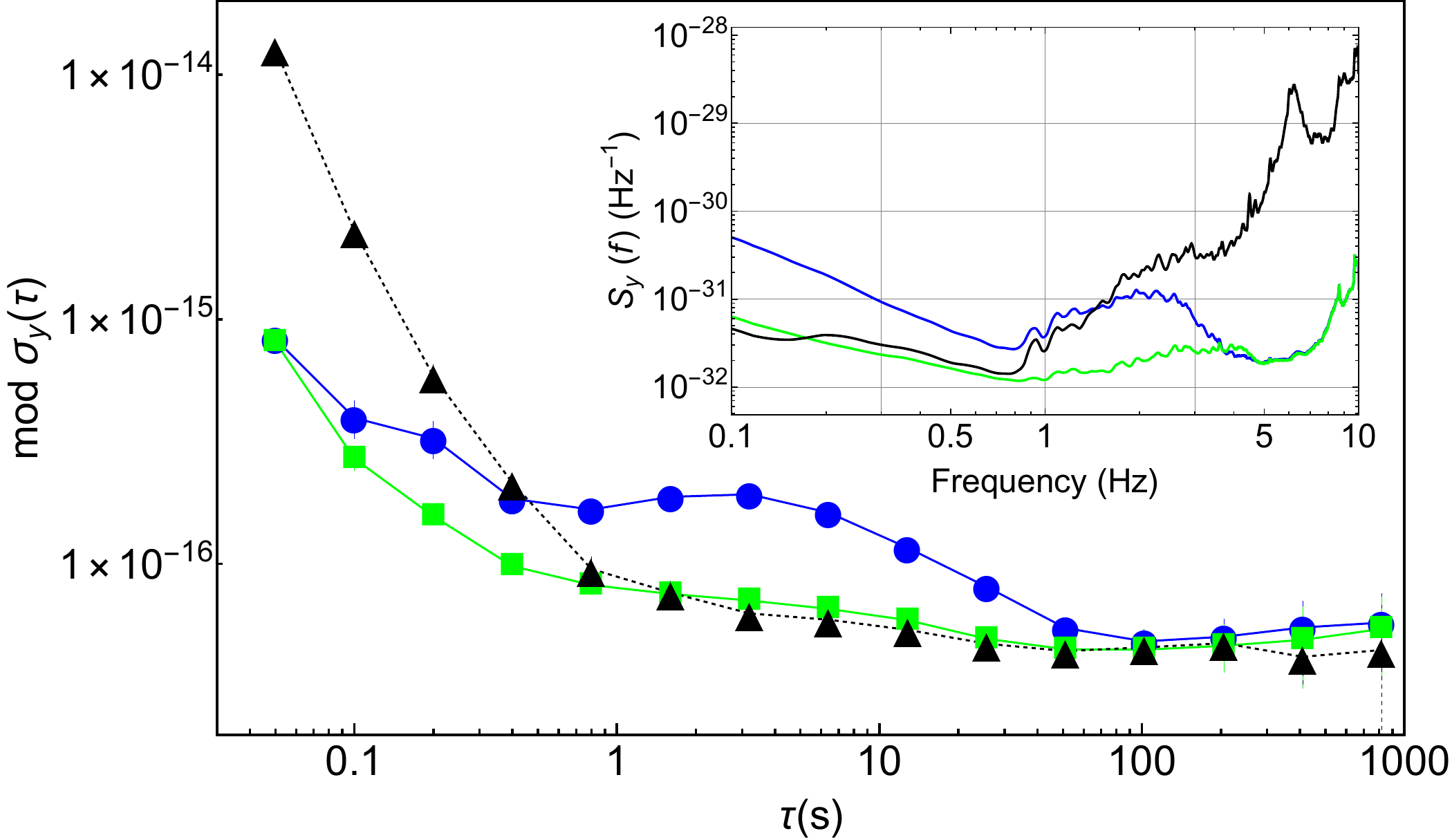}
	\caption{	\label{Fig:Vibration suppression}
			Fractional frequency instability of the cavity stabilized laser evaluated with TCH. The stability of the laser measured at $P_\mathrm{trans} = 52~\upmu$W without AVI is shown by black triangles and with AVI by blue circles. The stability obtained with feed forward correction of acceleration noise is shown by green squares. The power spectral density of fractional frequency noise $S_y (f)$ is shown in the inset with the same colors.   
	}
\end{figure}

As the cavity setup was not optimized for low vibration sensitivity we measure relatively high fractional frequency sensitivities to accelerations in the vertical direction of $1.0 \times 10^{-9}$/g, in the horizontal directions of $1.15 \times 10^{-9}$/g along the spacer length and $6.0 \times 10^{-11}$/g in the perpendicular direction where g is the standard gravity.
Thus the laser instability is rapidly increasing towards shorter averaging times (Fig. \ref{Fig:Vibration suppression} black). 
To suppress the high frequency vibrations we activate the AVI. 
In this case the vibration-induced frequency fluctuations for averaging times below 0.4~s are strongly suppressed, but at a cost of degraded stability between averaging times of 0.4~s and 100~s (Fig.~\ref{Fig:Vibration suppression} blue). 
The degradation is due to the noise of the internal sensors of the AVI systems that translates into extra tilt motion on the platforms \cite{bar25} which is also visible in the measured noise in the frequency domain (Fig.~\ref{Fig:Vibration suppression} inset). 

Here we investigate the prospects of a feed forward correction to mitigate the influence of accelerations. 
From the vertical acceleration measured by the seismometer and the horizontal accelerations derived from the tilt sensor, using the acceleration sensitivities of the cavity we first calculate the expected frequency fluctuation.
For a proof of principle demonstration, instead of applying the correction in real time, we do a post processing of our measured beat frequencies that are  synchronized to the measured acceleration data.  
We subtract the expected frequency fluctuation from the beat notes against the two silicon cavities and reevaluate the frequency stability with TCH. 
In this case the additional noise from the AVI platforms can be substantially removed in the processed data while maintaining a good stability at short and long averaging times as shown in Fig.~\ref{Fig:Vibration suppression}. 
With the feed forward corrections the noise is reduced between frequencies of 1~Hz and 4~Hz where the AVI is not suppressing vibrations and at frequencies below 1~Hz where the AVI is adding tilt noise as shown in the inset. 

\section{Conclusion and Outlook}

We have substantially reduced the gap in frequency stability between room temperature interferometers and more sophisticated cryogenic systems with a ULE cavity with fused silica substrates employing AlGaAs mirror coatings. 
Adding LED illumination of the mirrors we can make the cavity insensitive to laser power fluctuations at different intracavity power. 
After careful suppression of different technical noise contributions, the instability reaches $4.2\,(0.6) \times 10^{-17}$ which is one of the lowest instabilities for a room temperature cavity.
We give a comprehensive noise budget that well explains the observed stability over the range of investigated averaging times.
An important contribution in the budget is the birefringent noise first observed here at room temperature. At 1~s averaging time the birefringent noise is higher than the fundamental thermal noise of the system, which can be substantially reduced by polarization averaging techniques in the future \cite{yu23a, ked23}. 
The agreement between the noise budget and the stability precludes any other coating noise contribution larger than the low thermal noise of the cavity. This is critical for future room temperature based ultrastable lasers and gravitational wave telescopes employing AlGaAs coatings aiming at even better stability.    

At short averaging time, vibration noise is a major technical noise in state-of-the-art ultrastable lasers, including the current setup. We therefore demonstrated in this system that we can substantially remove the influence on the stabilized laser frequency due to vibrations and tilt motions using frequency feed forward method. 
Even when the sensitivities to accelerations of the current cavity are two orders of magnitude higher than the state-of-the-art systems, this method enables $10^{-17}$ instability for averaging times covering four orders of magnitude after real time frequency correction in the future.
This opens up a new possibility of researching materials with low thermal noise even when the mechanical properties are inferior. 
This method would be valuable for precise measurements during transportation of Fabry-Perot interferometers \cite{fre16}.

\section*{Acknowledgements}
We acknowledge support by the Project 20FUN08 NEXTLASERS, which has received funding from the EMPIR programme cofinanced by the Participating States and from the European Union’s Horizon 2020 Research and Innovation Programme, and by the Deutsche Forschungsgemeinschaft (DFG, German Research Foundation) under Germany’s Excellence Strategy–EX-2123 QuantumFrontiers (Project No. 390837967). This work is partially supported by the Max Planck-RIKEN-PTB Center for Time, Constants and Fundamental Symmetries. We thank Jun Ye and Garrett Cole for insightful discussions.

\bibliography{bib_short}

\begin{thebibliography}{36}%
\makeatletter
\providecommand \@ifxundefined [1]{%
 \@ifx{#1\undefined}
}%
\providecommand \@ifnum [1]{%
 \ifnum #1\expandafter \@firstoftwo
 \else \expandafter \@secondoftwo
 \fi
}%
\providecommand \@ifx [1]{%
 \ifx #1\expandafter \@firstoftwo
 \else \expandafter \@secondoftwo
 \fi
}%
\providecommand \natexlab [1]{#1}%
\providecommand \enquote  [1]{``#1''}%
\providecommand \bibnamefont  [1]{#1}%
\providecommand \bibfnamefont [1]{#1}%
\providecommand \citenamefont [1]{#1}%
\providecommand \href@noop [0]{\@secondoftwo}%
\providecommand \href [0]{\begingroup \@sanitize@url \@href}%
\providecommand \@href[1]{\@@startlink{#1}\@@href}%
\providecommand \@@href[1]{\endgroup#1\@@endlink}%
\providecommand \@sanitize@url [0]{\catcode `\\12\catcode `\$12\catcode
  `\&12\catcode `\#12\catcode `\^12\catcode `\_12\catcode `\%12\relax}%
\providecommand \@@startlink[1]{}%
\providecommand \@@endlink[0]{}%
\providecommand \url  [0]{\begingroup\@sanitize@url \@url }%
\providecommand \@url [1]{\endgroup\@href {#1}{\urlprefix }}%
\providecommand \urlprefix  [0]{URL }%
\providecommand \Eprint [0]{\href }%
\providecommand \doibase [0]{https://doi.org/}%
\providecommand \selectlanguage [0]{\@gobble}%
\providecommand \bibinfo  [0]{\@secondoftwo}%
\providecommand \bibfield  [0]{\@secondoftwo}%
\providecommand \translation [1]{[#1]}%
\providecommand \BibitemOpen [0]{}%
\providecommand \bibitemStop [0]{}%
\providecommand \bibitemNoStop [0]{.\EOS\space}%
\providecommand \EOS [0]{\spacefactor3000\relax}%
\providecommand \BibitemShut  [1]{\csname bibitem#1\endcsname}%
\let\auto@bib@innerbib\@empty
\bibitem [{\citenamefont {Ludlow}\ \emph {et~al.}(2015)\citenamefont {Ludlow},
  \citenamefont {Boyd}, \citenamefont {Ye}, \citenamefont {Peik},\ and\
  \citenamefont {Schmidt}}]{lud15}%
  \BibitemOpen
  \bibfield  {author} {\bibinfo {author} {\bibfnamefont {A.~D.}\ \bibnamefont
  {Ludlow}}, \bibinfo {author} {\bibfnamefont {M.~M.}\ \bibnamefont {Boyd}},
  \bibinfo {author} {\bibfnamefont {J.}~\bibnamefont {Ye}}, \bibinfo {author}
  {\bibfnamefont {E.}~\bibnamefont {Peik}},\ and\ \bibinfo {author}
  {\bibfnamefont {P.~O.}\ \bibnamefont {Schmidt}},\ }\bibfield  {title}
  {\bibinfo {title} {Optical atomic clocks},\ }\href
  {https://doi.org/10.1103/RevModPhys.87.637} {\bibfield  {journal} {\bibinfo
  {journal} {Rev. Mod. Phys.}\ }\textbf {\bibinfo {volume} {87}},\ \bibinfo
  {pages} {637} (\bibinfo {year} {2015})}\BibitemShut {NoStop}%
\bibitem [{\citenamefont {Fortier}\ \emph {et~al.}(2026)\citenamefont
  {Fortier}, \citenamefont {Luiten},\ and\ \citenamefont {Margolis}}]{for26}%
  \BibitemOpen
  \bibfield  {author} {\bibinfo {author} {\bibfnamefont {T.~M.}\ \bibnamefont
  {Fortier}}, \bibinfo {author} {\bibfnamefont {A.~N.}\ \bibnamefont
  {Luiten}},\ and\ \bibinfo {author} {\bibfnamefont {H.~S.}\ \bibnamefont
  {Margolis}},\ }\bibfield  {title} {\bibinfo {title} {Optical atomic clocks:
  defining the future of time and frequency metrology},\ }\href
  {https://doi.org/10.1364/OPTICA.575770} {\bibfield  {journal} {\bibinfo
  {journal} {Optica}\ }\textbf {\bibinfo {volume} {13}},\ \bibinfo {pages}
  {143} (\bibinfo {year} {2026})}\BibitemShut {NoStop}%
\bibitem [{\citenamefont {Wiens}\ \emph {et~al.}(2016)\citenamefont {Wiens},
  \citenamefont {Nevsky},\ and\ \citenamefont {Schiller}}]{wie16}%
  \BibitemOpen
  \bibfield  {author} {\bibinfo {author} {\bibfnamefont {E.}~\bibnamefont
  {Wiens}}, \bibinfo {author} {\bibfnamefont {A.~Y.}\ \bibnamefont {Nevsky}},\
  and\ \bibinfo {author} {\bibfnamefont {S.}~\bibnamefont {Schiller}},\
  }\bibfield  {title} {\bibinfo {title} {Resonator with ultrahigh length
  stability as a probe for equivalence-principle-violating physics},\ }\href
  {https://doi.org/10.1103/PhysRevLett.117.271102} {\bibfield  {journal}
  {\bibinfo  {journal} {Phys. Rev. Lett.}\ }\textbf {\bibinfo {volume} {117}},\
  \bibinfo {pages} {271102} (\bibinfo {year} {2016})}\BibitemShut {NoStop}%
\bibitem [{\citenamefont {Kennedy}\ \emph {et~al.}(2020)\citenamefont
  {Kennedy}, \citenamefont {Oelker}, \citenamefont {Robinson}, \citenamefont
  {Bothwell}, \citenamefont {Kedar}, \citenamefont {Milner}, \citenamefont
  {Marti}, \citenamefont {Derevianko},\ and\ \citenamefont {Ye}}]{ken20}%
  \BibitemOpen
  \bibfield  {author} {\bibinfo {author} {\bibfnamefont {C.~J.}\ \bibnamefont
  {Kennedy}}, \bibinfo {author} {\bibfnamefont {E.}~\bibnamefont {Oelker}},
  \bibinfo {author} {\bibfnamefont {J.~M.}\ \bibnamefont {Robinson}}, \bibinfo
  {author} {\bibfnamefont {T.}~\bibnamefont {Bothwell}}, \bibinfo {author}
  {\bibfnamefont {D.}~\bibnamefont {Kedar}}, \bibinfo {author} {\bibfnamefont
  {W.~R.}\ \bibnamefont {Milner}}, \bibinfo {author} {\bibfnamefont {G.~E.}\
  \bibnamefont {Marti}}, \bibinfo {author} {\bibfnamefont {A.}~\bibnamefont
  {Derevianko}},\ and\ \bibinfo {author} {\bibfnamefont {J.}~\bibnamefont
  {Ye}},\ }\bibfield  {title} {\bibinfo {title} {Precision metrology meets
  cosmology: {I}mproved constraints on ultralight dark matter from atom-cavity
  frequency comparisons},\ }\href
  {https://doi.org/10.1103/PhysRevLett.125.201302} {\bibfield  {journal}
  {\bibinfo  {journal} {Phys. Rev. Lett.}\ }\textbf {\bibinfo {volume} {125}},\
  \bibinfo {pages} {201302} (\bibinfo {year} {2020})}\BibitemShut {NoStop}%
\bibitem [{\citenamefont {Xie}\ \emph {et~al.}(2017)\citenamefont {Xie},
  \citenamefont {Bouchand}, \citenamefont {Nicolodi}, \citenamefont {Giunta},
  \citenamefont {H\"ansel}, \citenamefont {Lezius}, \citenamefont {Joshi},
  \citenamefont {Datta}, \citenamefont {Alexandre}, \citenamefont {Lours},
  \citenamefont {Tremblin}, \citenamefont {Santarelli}, \citenamefont
  {Holzwarth},\ and\ \citenamefont {Le~Coq}}]{xie17}%
  \BibitemOpen
  \bibfield  {author} {\bibinfo {author} {\bibfnamefont {X.}~\bibnamefont
  {Xie}}, \bibinfo {author} {\bibfnamefont {R.}~\bibnamefont {Bouchand}},
  \bibinfo {author} {\bibfnamefont {D.}~\bibnamefont {Nicolodi}}, \bibinfo
  {author} {\bibfnamefont {M.}~\bibnamefont {Giunta}}, \bibinfo {author}
  {\bibfnamefont {W.}~\bibnamefont {H\"ansel}}, \bibinfo {author}
  {\bibfnamefont {M.}~\bibnamefont {Lezius}}, \bibinfo {author} {\bibfnamefont
  {A.}~\bibnamefont {Joshi}}, \bibinfo {author} {\bibfnamefont
  {S.}~\bibnamefont {Datta}}, \bibinfo {author} {\bibfnamefont
  {C.}~\bibnamefont {Alexandre}}, \bibinfo {author} {\bibfnamefont
  {M.}~\bibnamefont {Lours}}, \bibinfo {author} {\bibfnamefont {P.-A.}\
  \bibnamefont {Tremblin}}, \bibinfo {author} {\bibfnamefont {G.}~\bibnamefont
  {Santarelli}}, \bibinfo {author} {\bibfnamefont {R.}~\bibnamefont
  {Holzwarth}},\ and\ \bibinfo {author} {\bibfnamefont {Y.}~\bibnamefont
  {Le~Coq}},\ }\bibfield  {title} {\bibinfo {title} {Photonic microwave signals
  with zeptosecond level absolute timing noise},\ }\href
  {https://doi.org/10.1038/nphoton.2016.215} {\bibfield  {journal} {\bibinfo
  {journal} {Nature Photonics}\ }\textbf {\bibinfo {volume} {11}},\ \bibinfo
  {pages} {44} (\bibinfo {year} {2017})}\BibitemShut {NoStop}%
\bibitem [{\citenamefont {Attar}\ \emph {et~al.}(2024)\citenamefont {Attar},
  \citenamefont {Timmers}, \citenamefont {Sodergren}, \citenamefont {Smith},
  \citenamefont {Barnes}, \citenamefont {Phillips}, \citenamefont {Vogel},\
  and\ \citenamefont {Knabe}}]{att24}%
  \BibitemOpen
  \bibfield  {author} {\bibinfo {author} {\bibfnamefont {A.}~\bibnamefont
  {Attar}}, \bibinfo {author} {\bibfnamefont {H.}~\bibnamefont {Timmers}},
  \bibinfo {author} {\bibfnamefont {B.}~\bibnamefont {Sodergren}}, \bibinfo
  {author} {\bibfnamefont {C.}~\bibnamefont {Smith}}, \bibinfo {author}
  {\bibfnamefont {E.}~\bibnamefont {Barnes}}, \bibinfo {author} {\bibfnamefont
  {N.}~\bibnamefont {Phillips}}, \bibinfo {author} {\bibfnamefont
  {K.}~\bibnamefont {Vogel}},\ and\ \bibinfo {author} {\bibfnamefont
  {K.}~\bibnamefont {Knabe}},\ }\bibfield  {title} {\bibinfo {title} {Ultra-low
  phase noise microwave generation using photonics for enhanced radar
  applications},\ }\href {https://doi.org/10.1117/12.3013763} {\bibfield
  {journal} {\bibinfo  {journal} {Proc. SPIE}\ }\textbf {\bibinfo {volume}
  {13048}},\ \bibinfo {pages} {130480O1} (\bibinfo {year} {2024})}\BibitemShut
  {NoStop}%
\bibitem [{\citenamefont {Meiser}\ \emph {et~al.}(2009)\citenamefont {Meiser},
  \citenamefont {Ye}, \citenamefont {Carlson},\ and\ \citenamefont
  {Holland}}]{mei09}%
  \BibitemOpen
  \bibfield  {author} {\bibinfo {author} {\bibfnamefont {D.}~\bibnamefont
  {Meiser}}, \bibinfo {author} {\bibfnamefont {J.}~\bibnamefont {Ye}}, \bibinfo
  {author} {\bibfnamefont {D.~R.}\ \bibnamefont {Carlson}},\ and\ \bibinfo
  {author} {\bibfnamefont {M.~J.}\ \bibnamefont {Holland}},\ }\bibfield
  {title} {\bibinfo {title} {Prospects for a millihertz-linewidth laser},\
  }\href {https://doi.org/10.1103/PhysRevLett.102.163601} {\bibfield  {journal}
  {\bibinfo  {journal} {Phys. Rev. Lett.}\ }\textbf {\bibinfo {volume} {102}},\
  \bibinfo {pages} {163601} (\bibinfo {year} {2009})}\BibitemShut {NoStop}%
\bibitem [{\citenamefont {Olson}\ \emph {et~al.}(2019)\citenamefont {Olson},
  \citenamefont {Fox}, \citenamefont {Fortier}, \citenamefont {Sheerin},
  \citenamefont {Brown}, \citenamefont {Leopardi}, \citenamefont {Stoner},
  \citenamefont {Oates},\ and\ \citenamefont {Ludlow}}]{ols19}%
  \BibitemOpen
  \bibfield  {author} {\bibinfo {author} {\bibfnamefont {J.}~\bibnamefont
  {Olson}}, \bibinfo {author} {\bibfnamefont {R.~W.}\ \bibnamefont {Fox}},
  \bibinfo {author} {\bibfnamefont {T.~M.}\ \bibnamefont {Fortier}}, \bibinfo
  {author} {\bibfnamefont {T.~F.}\ \bibnamefont {Sheerin}}, \bibinfo {author}
  {\bibfnamefont {R.~C.}\ \bibnamefont {Brown}}, \bibinfo {author}
  {\bibfnamefont {H.}~\bibnamefont {Leopardi}}, \bibinfo {author}
  {\bibfnamefont {R.~E.}\ \bibnamefont {Stoner}}, \bibinfo {author}
  {\bibfnamefont {C.~W.}\ \bibnamefont {Oates}},\ and\ \bibinfo {author}
  {\bibfnamefont {A.~D.}\ \bibnamefont {Ludlow}},\ }\bibfield  {title}
  {\bibinfo {title} {Ramsey-{B}ord\'e matter-wave interferometry for laser
  frequency stabilization at $10^{-16}$ frequency instability and below},\
  }\href {https://doi.org/10.1103/PhysRevLett.123.07320} {\bibfield  {journal}
  {\bibinfo  {journal} {Phys. Rev. Lett.}\ }\textbf {\bibinfo {volume} {123}},\
  \bibinfo {pages} {073202} (\bibinfo {year} {2019})}\BibitemShut {NoStop}%
\bibitem [{\citenamefont {Lin}\ \emph {et~al.}(2024)\citenamefont {Lin},
  \citenamefont {Hartman}, \citenamefont {Pointard}, \citenamefont {Le~Targat},
  \citenamefont {Goldner}, \citenamefont {Seidelin}, \citenamefont {Fang},\
  and\ \citenamefont {Le~Coq}}]{lin24}%
  \BibitemOpen
  \bibfield  {author} {\bibinfo {author} {\bibfnamefont {X.}~\bibnamefont
  {Lin}}, \bibinfo {author} {\bibfnamefont {M.~T.}\ \bibnamefont {Hartman}},
  \bibinfo {author} {\bibfnamefont {B.}~\bibnamefont {Pointard}}, \bibinfo
  {author} {\bibfnamefont {R.}~\bibnamefont {Le~Targat}}, \bibinfo {author}
  {\bibfnamefont {P.}~\bibnamefont {Goldner}}, \bibinfo {author} {\bibfnamefont
  {S.}~\bibnamefont {Seidelin}}, \bibinfo {author} {\bibfnamefont
  {B.}~\bibnamefont {Fang}},\ and\ \bibinfo {author} {\bibfnamefont
  {Y.}~\bibnamefont {Le~Coq}},\ }\bibfield  {title} {\bibinfo {title}
  {Anomalous subkelvin thermal frequency shifts of ultranarrow linewidth solid
  state emitters},\ }\href {https://doi.org/10.1103/PhysRevLett.133.183803}
  {\bibfield  {journal} {\bibinfo  {journal} {Phys. Rev. Lett.}\ }\textbf
  {\bibinfo {volume} {133}},\ \bibinfo {pages} {183803} (\bibinfo {year}
  {2024})}\BibitemShut {NoStop}%
\bibitem [{\citenamefont {Numata}\ \emph {et~al.}(2004)\citenamefont {Numata},
  \citenamefont {Kemery},\ and\ \citenamefont {Camp}}]{num04}%
  \BibitemOpen
  \bibfield  {author} {\bibinfo {author} {\bibfnamefont {K.}~\bibnamefont
  {Numata}}, \bibinfo {author} {\bibfnamefont {A.}~\bibnamefont {Kemery}},\
  and\ \bibinfo {author} {\bibfnamefont {J.}~\bibnamefont {Camp}},\ }\bibfield
  {title} {\bibinfo {title} {Thermal-noise limit in the frequency stabilization
  of lasers with rigid cavities},\ }\href
  {https://doi.org/10.1103/PhysRevLett.93.250602} {\bibfield  {journal}
  {\bibinfo  {journal} {Phys. Rev. Lett.}\ }\textbf {\bibinfo {volume} {93}},\
  \bibinfo {pages} {250602} (\bibinfo {year} {2004})}\BibitemShut {NoStop}%
\bibitem [{\citenamefont {Harry}\ \emph {et~al.}(2006)\citenamefont {Harry},
  \citenamefont {Armandula}, \citenamefont {Black}, \citenamefont {Crooks},
  \citenamefont {Cagnoli}, \citenamefont {Hough}, \citenamefont {Murray},
  \citenamefont {Reid}, \citenamefont {Rowan}, \citenamefont {Sneddon},
  \citenamefont {Fejer}, \citenamefont {Route},\ and\ \citenamefont
  {Penn}}]{har06b}%
  \BibitemOpen
  \bibfield  {author} {\bibinfo {author} {\bibfnamefont {G.~M.}\ \bibnamefont
  {Harry}}, \bibinfo {author} {\bibfnamefont {H.}~\bibnamefont {Armandula}},
  \bibinfo {author} {\bibfnamefont {E.}~\bibnamefont {Black}}, \bibinfo
  {author} {\bibfnamefont {D.~R.~M.}\ \bibnamefont {Crooks}}, \bibinfo {author}
  {\bibfnamefont {G.}~\bibnamefont {Cagnoli}}, \bibinfo {author} {\bibfnamefont
  {J.}~\bibnamefont {Hough}}, \bibinfo {author} {\bibfnamefont
  {P.}~\bibnamefont {Murray}}, \bibinfo {author} {\bibfnamefont
  {S.}~\bibnamefont {Reid}}, \bibinfo {author} {\bibfnamefont {S.}~\bibnamefont
  {Rowan}}, \bibinfo {author} {\bibfnamefont {P.}~\bibnamefont {Sneddon}},
  \bibinfo {author} {\bibfnamefont {M.~M.}\ \bibnamefont {Fejer}}, \bibinfo
  {author} {\bibfnamefont {R.}~\bibnamefont {Route}},\ and\ \bibinfo {author}
  {\bibfnamefont {S.~D.}\ \bibnamefont {Penn}},\ }\bibfield  {title} {\bibinfo
  {title} {Thermal noise from optical coatings in gravitational wave
  detectors},\ }\href {https://doi.org/10.1364/AO.45.001569} {\bibfield
  {journal} {\bibinfo  {journal} {Appl. Opt.}\ }\textbf {\bibinfo {volume}
  {45}},\ \bibinfo {pages} {1569} (\bibinfo {year} {2006})}\BibitemShut
  {NoStop}%
\bibitem [{\citenamefont {Cole}\ \emph {et~al.}(2013)\citenamefont {Cole},
  \citenamefont {Zhang}, \citenamefont {Martin}, \citenamefont {Ye},\ and\
  \citenamefont {Aspelmeyer}}]{col13}%
  \BibitemOpen
  \bibfield  {author} {\bibinfo {author} {\bibfnamefont {G.~D.}\ \bibnamefont
  {Cole}}, \bibinfo {author} {\bibfnamefont {W.}~\bibnamefont {Zhang}},
  \bibinfo {author} {\bibfnamefont {M.~J.}\ \bibnamefont {Martin}}, \bibinfo
  {author} {\bibfnamefont {J.}~\bibnamefont {Ye}},\ and\ \bibinfo {author}
  {\bibfnamefont {M.}~\bibnamefont {Aspelmeyer}},\ }\bibfield  {title}
  {\bibinfo {title} {Tenfold reduction of {B}rownian noise in optical
  interferometry},\ }\href {https://doi.org/10.1038/NPHOTON.2013.174}
  {\bibfield  {journal} {\bibinfo  {journal} {Nature Photonics}\ }\textbf
  {\bibinfo {volume} {7}},\ \bibinfo {pages} {644} (\bibinfo {year}
  {2013})}\BibitemShut {NoStop}%
\bibitem [{\citenamefont {Yu}\ \emph {et~al.}(2023)\citenamefont {Yu},
  \citenamefont {H\"afner}, \citenamefont {Legero}, \citenamefont {Herbers},
  \citenamefont {Nicolodi}, \citenamefont {Ma}, \citenamefont {Riehle},
  \citenamefont {Sterr}, \citenamefont {Kedar}, \citenamefont {Robinson},
  \citenamefont {Oelker},\ and\ \citenamefont {Ye}}]{yu23a}%
  \BibitemOpen
  \bibfield  {author} {\bibinfo {author} {\bibfnamefont {J.}~\bibnamefont
  {Yu}}, \bibinfo {author} {\bibfnamefont {S.}~\bibnamefont {H\"afner}},
  \bibinfo {author} {\bibfnamefont {T.}~\bibnamefont {Legero}}, \bibinfo
  {author} {\bibfnamefont {S.}~\bibnamefont {Herbers}}, \bibinfo {author}
  {\bibfnamefont {D.}~\bibnamefont {Nicolodi}}, \bibinfo {author}
  {\bibfnamefont {C.~Y.}\ \bibnamefont {Ma}}, \bibinfo {author} {\bibfnamefont
  {F.}~\bibnamefont {Riehle}}, \bibinfo {author} {\bibfnamefont
  {U.}~\bibnamefont {Sterr}}, \bibinfo {author} {\bibfnamefont
  {D.}~\bibnamefont {Kedar}}, \bibinfo {author} {\bibfnamefont {J.~M.}\
  \bibnamefont {Robinson}}, \bibinfo {author} {\bibfnamefont {E.}~\bibnamefont
  {Oelker}},\ and\ \bibinfo {author} {\bibfnamefont {J.}~\bibnamefont {Ye}},\
  }\bibfield  {title} {\bibinfo {title} {Excess noise and photoinduced effects
  in highly reflective crystalline mirror coatings},\ }\href
  {https://doi.org/10.1103/PhysRevX.13.041002} {\bibfield  {journal} {\bibinfo
  {journal} {Phys. Rev. X}\ }\textbf {\bibinfo {volume} {13}},\ \bibinfo
  {pages} {041002} (\bibinfo {year} {2023})}\BibitemShut {NoStop}%
\bibitem [{\citenamefont {Lee}\ \emph {et~al.}(2026)\citenamefont {Lee},
  \citenamefont {Hu}, \citenamefont {Lewis}, \citenamefont {Aeppli},
  \citenamefont {Kim}, \citenamefont {Yao}, \citenamefont {Legero},
  \citenamefont {Nicolodi}, \citenamefont {Riehle}, \citenamefont {Sterr},\
  and\ \citenamefont {Ye}}]{lee26}%
  \BibitemOpen
  \bibfield  {author} {\bibinfo {author} {\bibfnamefont {D.}~\bibnamefont
  {Lee}}, \bibinfo {author} {\bibfnamefont {Z.~Z.}\ \bibnamefont {Hu}},
  \bibinfo {author} {\bibfnamefont {B.}~\bibnamefont {Lewis}}, \bibinfo
  {author} {\bibfnamefont {A.}~\bibnamefont {Aeppli}}, \bibinfo {author}
  {\bibfnamefont {K.}~\bibnamefont {Kim}}, \bibinfo {author} {\bibfnamefont
  {Z.}~\bibnamefont {Yao}}, \bibinfo {author} {\bibfnamefont {T.}~\bibnamefont
  {Legero}}, \bibinfo {author} {\bibfnamefont {D.}~\bibnamefont {Nicolodi}},
  \bibinfo {author} {\bibfnamefont {F.}~\bibnamefont {Riehle}}, \bibinfo
  {author} {\bibfnamefont {U.}~\bibnamefont {Sterr}},\ and\ \bibinfo {author}
  {\bibfnamefont {J.}~\bibnamefont {Ye}},\ }\bibfield  {title} {\bibinfo
  {title} {Frequency stability of $2.5\times 10^{-17}$ from a {Si} cavity with
  {AlGaAs} crystalline mirrors},\ }\href {https://doi.org/10.1103/zgrm-cjbb}
  {\bibfield  {journal} {\bibinfo  {journal} {Phys. Rev. Lett.}\ }\textbf
  {\bibinfo {volume} {136}},\ \bibinfo {pages} {033801} (\bibinfo {year}
  {2026})}\BibitemShut {NoStop}%
\bibitem [{\citenamefont {H{\"a}fner}\ \emph {et~al.}(2015)\citenamefont
  {H{\"a}fner}, \citenamefont {Falke}, \citenamefont {Grebing}, \citenamefont
  {Vogt}, \citenamefont {Legero}, \citenamefont {Merimaa}, \citenamefont
  {Lisdat},\ and\ \citenamefont {Sterr}}]{hae15a}%
  \BibitemOpen
  \bibfield  {author} {\bibinfo {author} {\bibfnamefont {S.}~\bibnamefont
  {H{\"a}fner}}, \bibinfo {author} {\bibfnamefont {S.}~\bibnamefont {Falke}},
  \bibinfo {author} {\bibfnamefont {C.}~\bibnamefont {Grebing}}, \bibinfo
  {author} {\bibfnamefont {S.}~\bibnamefont {Vogt}}, \bibinfo {author}
  {\bibfnamefont {T.}~\bibnamefont {Legero}}, \bibinfo {author} {\bibfnamefont
  {M.}~\bibnamefont {Merimaa}}, \bibinfo {author} {\bibfnamefont
  {C.}~\bibnamefont {Lisdat}},\ and\ \bibinfo {author} {\bibfnamefont
  {U.}~\bibnamefont {Sterr}},\ }\bibfield  {title} {\bibinfo {title} {$8 \times
  10^{-17}$ fractional laser frequency instability with a long room-temperature
  cavity},\ }\href {https://doi.org/10.1364/OL.40.002112} {\bibfield  {journal}
  {\bibinfo  {journal} {Opt. Lett.}\ }\textbf {\bibinfo {volume} {40}},\
  \bibinfo {pages} {2112} (\bibinfo {year} {2015})}\BibitemShut {NoStop}%
\bibitem [{\citenamefont {Schioppo}\ \emph {et~al.}(2022)\citenamefont
  {Schioppo}, \citenamefont {Kronj\"ager}, \citenamefont {Silva}, \citenamefont
  {Ilieva}, \citenamefont {Paterson}, \citenamefont {Baynham}, \citenamefont
  {Bowden}, \citenamefont {Hill}, \citenamefont {Hobson}, \citenamefont
  {Vianello}, \citenamefont {Dovale-\'Alvarez}, \citenamefont {Williams},
  \citenamefont {Marra}, \citenamefont {Margolis}, \citenamefont {Amy-Klein},
  \citenamefont {Lopez}, \citenamefont {Cantin}, \citenamefont
  {\'Alvarez-Mart\'inez}, \citenamefont {Le~Targat}, \citenamefont {Pottie},
  \citenamefont {Quintin}, \citenamefont {Legero}, \citenamefont {H\"afner},
  \citenamefont {Sterr}, \citenamefont {Schwarz}, \citenamefont {D\"orscher},
  \citenamefont {Lisdat}, \citenamefont {Koke}, \citenamefont {Kuhl},
  \citenamefont {Waterholter}, \citenamefont {Benkler},\ and\ \citenamefont
  {Grosche}}]{sch22a}%
  \BibitemOpen
  \bibfield  {author} {\bibinfo {author} {\bibfnamefont {M.}~\bibnamefont
  {Schioppo}}, \bibinfo {author} {\bibfnamefont {J.}~\bibnamefont
  {Kronj\"ager}}, \bibinfo {author} {\bibfnamefont {A.}~\bibnamefont {Silva}},
  \bibinfo {author} {\bibfnamefont {R.}~\bibnamefont {Ilieva}}, \bibinfo
  {author} {\bibfnamefont {J.~W.}\ \bibnamefont {Paterson}}, \bibinfo {author}
  {\bibfnamefont {C.~F.~A.}\ \bibnamefont {Baynham}}, \bibinfo {author}
  {\bibfnamefont {W.}~\bibnamefont {Bowden}}, \bibinfo {author} {\bibfnamefont
  {I.~R.}\ \bibnamefont {Hill}}, \bibinfo {author} {\bibfnamefont
  {R.}~\bibnamefont {Hobson}}, \bibinfo {author} {\bibfnamefont
  {A.}~\bibnamefont {Vianello}}, \bibinfo {author} {\bibfnamefont
  {M.}~\bibnamefont {Dovale-\'Alvarez}}, \bibinfo {author} {\bibfnamefont
  {R.~A.}\ \bibnamefont {Williams}}, \bibinfo {author} {\bibfnamefont
  {G.}~\bibnamefont {Marra}}, \bibinfo {author} {\bibfnamefont {H.~S.}\
  \bibnamefont {Margolis}}, \bibinfo {author} {\bibfnamefont {A.}~\bibnamefont
  {Amy-Klein}}, \bibinfo {author} {\bibfnamefont {O.}~\bibnamefont {Lopez}},
  \bibinfo {author} {\bibfnamefont {E.}~\bibnamefont {Cantin}}, \bibinfo
  {author} {\bibfnamefont {H.}~\bibnamefont {\'Alvarez-Mart\'inez}}, \bibinfo
  {author} {\bibfnamefont {R.}~\bibnamefont {Le~Targat}}, \bibinfo {author}
  {\bibfnamefont {P.~E.}\ \bibnamefont {Pottie}}, \bibinfo {author}
  {\bibfnamefont {N.}~\bibnamefont {Quintin}}, \bibinfo {author} {\bibfnamefont
  {T.}~\bibnamefont {Legero}}, \bibinfo {author} {\bibfnamefont
  {S.}~\bibnamefont {H\"afner}}, \bibinfo {author} {\bibfnamefont
  {U.}~\bibnamefont {Sterr}}, \bibinfo {author} {\bibfnamefont
  {R.}~\bibnamefont {Schwarz}}, \bibinfo {author} {\bibfnamefont
  {S.}~\bibnamefont {D\"orscher}}, \bibinfo {author} {\bibfnamefont
  {C.}~\bibnamefont {Lisdat}}, \bibinfo {author} {\bibfnamefont
  {S.}~\bibnamefont {Koke}}, \bibinfo {author} {\bibfnamefont {A.}~\bibnamefont
  {Kuhl}}, \bibinfo {author} {\bibfnamefont {T.}~\bibnamefont {Waterholter}},
  \bibinfo {author} {\bibfnamefont {E.}~\bibnamefont {Benkler}},\ and\ \bibinfo
  {author} {\bibfnamefont {G.}~\bibnamefont {Grosche}},\ }\bibfield  {title}
  {\bibinfo {title} {Comparing ultrastable lasers at $7\times 10^{-17}$
  fractional frequency instability through a 2220~km optical fibre network},\
  }\href {https://doi.org/10.1038/s41467-021-27884-3} {\bibfield  {journal}
  {\bibinfo  {journal} {Nature Commun.}\ }\textbf {\bibinfo {volume} {13}},\
  \bibinfo {pages} {212} (\bibinfo {year} {2022})}\BibitemShut {NoStop}%
\bibitem [{\citenamefont {Ma}\ \emph {et~al.}(2024)\citenamefont {Ma},
  \citenamefont {Yu}, \citenamefont {Legero}, \citenamefont {Herbers},
  \citenamefont {Nicolodi}, \citenamefont {Kempkes}, \citenamefont {Riehle},
  \citenamefont {Kedar}, \citenamefont {Robinson}, \citenamefont {Ye},\ and\
  \citenamefont {Sterr}}]{ma24a}%
  \BibitemOpen
  \bibfield  {author} {\bibinfo {author} {\bibfnamefont {C.~Y.}\ \bibnamefont
  {Ma}}, \bibinfo {author} {\bibfnamefont {J.}~\bibnamefont {Yu}}, \bibinfo
  {author} {\bibfnamefont {T.}~\bibnamefont {Legero}}, \bibinfo {author}
  {\bibfnamefont {S.}~\bibnamefont {Herbers}}, \bibinfo {author} {\bibfnamefont
  {D.}~\bibnamefont {Nicolodi}}, \bibinfo {author} {\bibfnamefont
  {M.}~\bibnamefont {Kempkes}}, \bibinfo {author} {\bibfnamefont
  {F.}~\bibnamefont {Riehle}}, \bibinfo {author} {\bibfnamefont
  {D.}~\bibnamefont {Kedar}}, \bibinfo {author} {\bibfnamefont {J.~M.}\
  \bibnamefont {Robinson}}, \bibinfo {author} {\bibfnamefont {J.}~\bibnamefont
  {Ye}},\ and\ \bibinfo {author} {\bibfnamefont {U.}~\bibnamefont {Sterr}},\
  }\bibfield  {title} {\bibinfo {title} {Ultrastable lasers: investigations of
  crystalline mirrors and closed cycle cooling at 124~{K}},\ }\href
  {https://doi.org/10.1088/1742-6596/2889/1/012055} {\bibfield  {journal}
  {\bibinfo  {journal} {J. Phys.: Conf. Ser.}\ }\textbf {\bibinfo {volume}
  {2889}},\ \bibinfo {pages} {012055} (\bibinfo {year} {2024})}\BibitemShut
  {NoStop}%
\bibitem [{\citenamefont {Parke}\ \emph {et~al.}(2026)\citenamefont {Parke},
  \citenamefont {Clulow}, \citenamefont {Huang}, \citenamefont {Kaur},
  \citenamefont {Karembera}, \citenamefont {Gaudron}, \citenamefont {Zhang},
  \citenamefont {Risaro}, \citenamefont {Tunesi}, \citenamefont {Bourne},\ and\
  \citenamefont {Schioppo}}]{par26}%
  \BibitemOpen
  \bibfield  {author} {\bibinfo {author} {\bibfnamefont {A.~L.}\ \bibnamefont
  {Parke}}, \bibinfo {author} {\bibfnamefont {E.}~\bibnamefont {Clulow}},
  \bibinfo {author} {\bibfnamefont {W.}~\bibnamefont {Huang}}, \bibinfo
  {author} {\bibfnamefont {N.}~\bibnamefont {Kaur}}, \bibinfo {author}
  {\bibfnamefont {R.}~\bibnamefont {Karembera}}, \bibinfo {author}
  {\bibfnamefont {J.-O.}\ \bibnamefont {Gaudron}}, \bibinfo {author}
  {\bibfnamefont {X.}~\bibnamefont {Zhang}}, \bibinfo {author} {\bibfnamefont
  {M.}~\bibnamefont {Risaro}}, \bibinfo {author} {\bibfnamefont
  {J.}~\bibnamefont {Tunesi}}, \bibinfo {author} {\bibfnamefont
  {H.}~\bibnamefont {Bourne}},\ and\ \bibinfo {author} {\bibfnamefont
  {M.}~\bibnamefont {Schioppo}},\ }\bibfield  {title} {\bibinfo {title} {Laser
  fractional frequency instability at $4 \times 10^{-17}$ with a room
  temperature optical reference cavity},\ }\href
  {https://doi.org/10.1364/OPTICA.591175} {\bibfield  {journal} {\bibinfo
  {journal} {Optica}\ }\textbf {\bibinfo {volume} {13}},\ \bibinfo {pages}
  {1190} (\bibinfo {year} {2026})}\BibitemShut {NoStop}%
\bibitem [{\citenamefont {Kedar}\ \emph {et~al.}(2023)\citenamefont {Kedar},
  \citenamefont {Yu}, \citenamefont {Oelker}, \citenamefont {Staron},
  \citenamefont {Milner}, \citenamefont {Robinson}, \citenamefont {Legero},
  \citenamefont {Riehle}, \citenamefont {Sterr},\ and\ \citenamefont
  {Ye}}]{ked23}%
  \BibitemOpen
  \bibfield  {author} {\bibinfo {author} {\bibfnamefont {D.}~\bibnamefont
  {Kedar}}, \bibinfo {author} {\bibfnamefont {J.}~\bibnamefont {Yu}}, \bibinfo
  {author} {\bibfnamefont {E.}~\bibnamefont {Oelker}}, \bibinfo {author}
  {\bibfnamefont {A.}~\bibnamefont {Staron}}, \bibinfo {author} {\bibfnamefont
  {W.~R.}\ \bibnamefont {Milner}}, \bibinfo {author} {\bibfnamefont {J.~M.}\
  \bibnamefont {Robinson}}, \bibinfo {author} {\bibfnamefont {T.}~\bibnamefont
  {Legero}}, \bibinfo {author} {\bibfnamefont {F.}~\bibnamefont {Riehle}},
  \bibinfo {author} {\bibfnamefont {U.}~\bibnamefont {Sterr}},\ and\ \bibinfo
  {author} {\bibfnamefont {J.}~\bibnamefont {Ye}},\ }\bibfield  {title}
  {\bibinfo {title} {Frequency stability of cryogenic silicon cavities with
  semiconductor crystalline coatings},\ }\href
  {https://doi.org/10.1364/OPTICA.479462} {\bibfield  {journal} {\bibinfo
  {journal} {Optica}\ }\textbf {\bibinfo {volume} {10}},\ \bibinfo {pages}
  {464} (\bibinfo {year} {2023})}\BibitemShut {NoStop}%
\bibitem [{\citenamefont {Winkler}\ \emph {et~al.}(2021)\citenamefont
  {Winkler}, \citenamefont {Perner}, \citenamefont {Truong}, \citenamefont
  {Zhao}, \citenamefont {Bachmann}, \citenamefont {Mayer}, \citenamefont
  {Fellinger}, \citenamefont {Follman}, \citenamefont {Heu}, \citenamefont
  {Deutsch}, \citenamefont {Bailey}, \citenamefont {Peelaers}, \citenamefont
  {Puchegger}, \citenamefont {Fleisher}, \citenamefont {Cole},\ and\
  \citenamefont {Heckl}}]{win21}%
  \BibitemOpen
  \bibfield  {author} {\bibinfo {author} {\bibfnamefont {G.}~\bibnamefont
  {Winkler}}, \bibinfo {author} {\bibfnamefont {L.~W.}\ \bibnamefont {Perner}},
  \bibinfo {author} {\bibfnamefont {G.-W.}\ \bibnamefont {Truong}}, \bibinfo
  {author} {\bibfnamefont {G.}~\bibnamefont {Zhao}}, \bibinfo {author}
  {\bibfnamefont {D.}~\bibnamefont {Bachmann}}, \bibinfo {author}
  {\bibfnamefont {A.~S.}\ \bibnamefont {Mayer}}, \bibinfo {author}
  {\bibfnamefont {J.}~\bibnamefont {Fellinger}}, \bibinfo {author}
  {\bibfnamefont {D.}~\bibnamefont {Follman}}, \bibinfo {author} {\bibfnamefont
  {P.}~\bibnamefont {Heu}}, \bibinfo {author} {\bibfnamefont {C.}~\bibnamefont
  {Deutsch}}, \bibinfo {author} {\bibfnamefont {D.~M.}\ \bibnamefont {Bailey}},
  \bibinfo {author} {\bibfnamefont {H.}~\bibnamefont {Peelaers}}, \bibinfo
  {author} {\bibfnamefont {S.}~\bibnamefont {Puchegger}}, \bibinfo {author}
  {\bibfnamefont {A.~J.}\ \bibnamefont {Fleisher}}, \bibinfo {author}
  {\bibfnamefont {G.~D.}\ \bibnamefont {Cole}},\ and\ \bibinfo {author}
  {\bibfnamefont {O.~H.}\ \bibnamefont {Heckl}},\ }\bibfield  {title} {\bibinfo
  {title} {Mid-infrared interference coatings with excess optical loss below
  10~ppm},\ }\href {https://doi.org/10.1364/OPTICA.405938} {\bibfield
  {journal} {\bibinfo  {journal} {Optica}\ }\textbf {\bibinfo {volume} {8}},\
  \bibinfo {pages} {686} (\bibinfo {year} {2021})},\ \bibinfo {note} {also see
  erratum \cite{per24}}\BibitemShut {NoStop}%
\bibitem [{\citenamefont {Legero}\ \emph {et~al.}(2010)\citenamefont {Legero},
  \citenamefont {Kessler},\ and\ \citenamefont {Sterr}}]{leg10}%
  \BibitemOpen
  \bibfield  {author} {\bibinfo {author} {\bibfnamefont {T.}~\bibnamefont
  {Legero}}, \bibinfo {author} {\bibfnamefont {T.}~\bibnamefont {Kessler}},\
  and\ \bibinfo {author} {\bibfnamefont {U.}~\bibnamefont {Sterr}},\ }\bibfield
   {title} {\bibinfo {title} {Tuning the thermal expansion properties of
  optical reference cavities with fused silica mirrors},\ }\href
  {https://doi.org/10.1364/JOSAB.27.000914} {\bibfield  {journal} {\bibinfo
  {journal} {J. Opt. Soc. Am. B}\ }\textbf {\bibinfo {volume} {27}},\ \bibinfo
  {pages} {914} (\bibinfo {year} {2010})}\BibitemShut {NoStop}%
\bibitem [{Note1()}]{Note1}%
  \BibitemOpen
  \bibinfo {note} {The Table Stable Ltd. AVI 200-M}\BibitemShut {NoStop}%
\bibitem [{\citenamefont {Black}(2001)}]{bla01}%
  \BibitemOpen
  \bibfield  {author} {\bibinfo {author} {\bibfnamefont {E.~D.}\ \bibnamefont
  {Black}},\ }\bibfield  {title} {\bibinfo {title} {An introduction to
  {P}ound-{D}rever-{H}all laser frequency stabilization},\ }\href
  {https://doi.org/10.1119/1.1286663} {\bibfield  {journal} {\bibinfo
  {journal} {Am. J. Phys.}\ }\textbf {\bibinfo {volume} {69}},\ \bibinfo
  {pages} {79} (\bibinfo {year} {2001})}\BibitemShut {NoStop}%
\bibitem [{\citenamefont {Matei}\ \emph {et~al.}(2017)\citenamefont {Matei},
  \citenamefont {Legero}, \citenamefont {H\"afner}, \citenamefont {Grebing},
  \citenamefont {Weyrich}, \citenamefont {Zhang}, \citenamefont {Sonderhouse},
  \citenamefont {Robinson}, \citenamefont {Ye}, \citenamefont {Riehle},\ and\
  \citenamefont {Sterr}}]{mat17a}%
  \BibitemOpen
  \bibfield  {author} {\bibinfo {author} {\bibfnamefont {D.~G.}\ \bibnamefont
  {Matei}}, \bibinfo {author} {\bibfnamefont {T.}~\bibnamefont {Legero}},
  \bibinfo {author} {\bibfnamefont {S.}~\bibnamefont {H\"afner}}, \bibinfo
  {author} {\bibfnamefont {C.}~\bibnamefont {Grebing}}, \bibinfo {author}
  {\bibfnamefont {R.}~\bibnamefont {Weyrich}}, \bibinfo {author} {\bibfnamefont
  {W.}~\bibnamefont {Zhang}}, \bibinfo {author} {\bibfnamefont
  {L.}~\bibnamefont {Sonderhouse}}, \bibinfo {author} {\bibfnamefont {J.~M.}\
  \bibnamefont {Robinson}}, \bibinfo {author} {\bibfnamefont {J.}~\bibnamefont
  {Ye}}, \bibinfo {author} {\bibfnamefont {F.}~\bibnamefont {Riehle}},\ and\
  \bibinfo {author} {\bibfnamefont {U.}~\bibnamefont {Sterr}},\ }\bibfield
  {title} {\bibinfo {title} {$1.5~\mu$m lasers with sub-{10 mHz} linewidth},\
  }\href {https://doi.org/10.1103/PhysRevLett.118.263202} {\bibfield  {journal}
  {\bibinfo  {journal} {Phys. Rev. Lett.}\ }\textbf {\bibinfo {volume} {118}},\
  \bibinfo {pages} {263202} (\bibinfo {year} {2017})}\BibitemShut {NoStop}%
\bibitem [{\citenamefont {Zhu}\ \emph {et~al.}(2024)\citenamefont {Zhu},
  \citenamefont {Cui}, \citenamefont {Kong}, \citenamefont {Yu}, \citenamefont
  {Zhai}, \citenamefont {Zheng}, \citenamefont {Xie}, \citenamefont {Zhang},
  \citenamefont {Jiang}, \citenamefont {Zhang}, \citenamefont {Xu},
  \citenamefont {Dai}, \citenamefont {Chen},\ and\ \citenamefont
  {Pan}}]{zhu24}%
  \BibitemOpen
  \bibfield  {author} {\bibinfo {author} {\bibfnamefont {X.-Q.}\ \bibnamefont
  {Zhu}}, \bibinfo {author} {\bibfnamefont {X.-Y.}\ \bibnamefont {Cui}},
  \bibinfo {author} {\bibfnamefont {D.-Q.}\ \bibnamefont {Kong}}, \bibinfo
  {author} {\bibfnamefont {H.-W.}\ \bibnamefont {Yu}}, \bibinfo {author}
  {\bibfnamefont {X.-M.}\ \bibnamefont {Zhai}}, \bibinfo {author}
  {\bibfnamefont {M.-Y.}\ \bibnamefont {Zheng}}, \bibinfo {author}
  {\bibfnamefont {X.-P.}\ \bibnamefont {Xie}}, \bibinfo {author} {\bibfnamefont
  {Q.}~\bibnamefont {Zhang}}, \bibinfo {author} {\bibfnamefont
  {X.}~\bibnamefont {Jiang}}, \bibinfo {author} {\bibfnamefont {X.-B.}\
  \bibnamefont {Zhang}}, \bibinfo {author} {\bibfnamefont {P.}~\bibnamefont
  {Xu}}, \bibinfo {author} {\bibfnamefont {H.-N.}\ \bibnamefont {Dai}},
  \bibinfo {author} {\bibfnamefont {Y.-A.}\ \bibnamefont {Chen}},\ and\
  \bibinfo {author} {\bibfnamefont {J.-W.}\ \bibnamefont {Pan}},\ }\bibfield
  {title} {\bibinfo {title} {An ultrastable 1397-nm laser stabilized by a
  crystalline-coated room-temperature cavity},\ }\href
  {https://doi.org/10.1063/5.0200553} {\bibfield  {journal} {\bibinfo
  {journal} {Rev. Sci. Instrum.}\ }\textbf {\bibinfo {volume} {95}},\ \bibinfo
  {pages} {083002} (\bibinfo {year} {2024})}\BibitemShut {NoStop}%
\bibitem [{\citenamefont {Kraus}\ \emph {et~al.}(2025)\citenamefont {Kraus},
  \citenamefont {Herbers}, \citenamefont {Nauk}, \citenamefont {Sterr},
  \citenamefont {Lisdat},\ and\ \citenamefont {Schmidt}}]{kra25a}%
  \BibitemOpen
  \bibfield  {author} {\bibinfo {author} {\bibfnamefont {B.}~\bibnamefont
  {Kraus}}, \bibinfo {author} {\bibfnamefont {S.}~\bibnamefont {Herbers}},
  \bibinfo {author} {\bibfnamefont {C.}~\bibnamefont {Nauk}}, \bibinfo {author}
  {\bibfnamefont {U.}~\bibnamefont {Sterr}}, \bibinfo {author} {\bibfnamefont
  {C.}~\bibnamefont {Lisdat}},\ and\ \bibinfo {author} {\bibfnamefont {P.~O.}\
  \bibnamefont {Schmidt}},\ }\bibfield  {title} {\bibinfo {title} {Ultra-stable
  transportable ultraviolet clock laser using cancellation between
  photo-thermal and photo-birefringence noise},\ }\href
  {https://doi.org/10.1364/OL.544907} {\bibfield  {journal} {\bibinfo
  {journal} {Opt. Lett.}\ }\textbf {\bibinfo {volume} {50}},\ \bibinfo {pages}
  {658} (\bibinfo {year} {2025})}\BibitemShut {NoStop}%
\bibitem [{\citenamefont {Wu}\ \emph {et~al.}(2025)\citenamefont {Wu},
  \citenamefont {Goswami}, \citenamefont {Tanioka},\ and\ \citenamefont
  {Ballmer}}]{wu25c}%
  \BibitemOpen
  \bibfield  {author} {\bibinfo {author} {\bibfnamefont {B.}~\bibnamefont
  {Wu}}, \bibinfo {author} {\bibfnamefont {S.}~\bibnamefont {Goswami}},
  \bibinfo {author} {\bibfnamefont {S.}~\bibnamefont {Tanioka}},\ and\ \bibinfo
  {author} {\bibfnamefont {S.}~\bibnamefont {Ballmer}},\ }\href
  {https://doi.org/10.48550/arXiv.2512.00594} {\bibinfo {title} {Birefringence
  of {AlGaAs}/{GaA}s coatings under above-band-gap illumination, {GR} noise and
  photo-optic transfer function}},\ \bibinfo {howpublished} {arXiv:2512.00594
  [physics.ins-det]} (\bibinfo {year} {2025}),\ \bibinfo {note} {lIGO Document
  P2500676-v2}\BibitemShut {NoStop}%
\bibitem [{\citenamefont {Ma}\ \emph {et~al.}(2026)\citenamefont {Ma},
  \citenamefont {Yu}, \citenamefont {Legero}, \citenamefont {Herbers},
  \citenamefont {Nicolodi}, \citenamefont {Kempkes}, \citenamefont {Riehle},\
  and\ \citenamefont {Sterr}}]{ma26}%
  \BibitemOpen
  \bibfield  {author} {\bibinfo {author} {\bibfnamefont {C.~Y.}\ \bibnamefont
  {Ma}}, \bibinfo {author} {\bibfnamefont {J.}~\bibnamefont {Yu}}, \bibinfo
  {author} {\bibfnamefont {T.}~\bibnamefont {Legero}}, \bibinfo {author}
  {\bibfnamefont {S.}~\bibnamefont {Herbers}}, \bibinfo {author} {\bibfnamefont
  {D.}~\bibnamefont {Nicolodi}}, \bibinfo {author} {\bibfnamefont
  {M.}~\bibnamefont {Kempkes}}, \bibinfo {author} {\bibfnamefont
  {F.}~\bibnamefont {Riehle}},\ and\ \bibinfo {author} {\bibfnamefont
  {U.}~\bibnamefont {Sterr}},\ }\bibfield  {title} {\bibinfo {title}
  {Photo-birefringent effects in crystalline {AlGaAs} mirror coatings},\ }\href
  {https://doi.org/10.1364/OE.593424} {\bibfield  {journal} {\bibinfo
  {journal} {Opt. Express}\ }\textbf {\bibinfo {volume} {34}},\ \bibinfo
  {pages} {25456} (\bibinfo {year} {2026})}\BibitemShut {NoStop}%
\bibitem [{\citenamefont {Evans}\ \emph {et~al.}(2008)\citenamefont {Evans},
  \citenamefont {Ballmer}, \citenamefont {Fejer}, \citenamefont {Fritschel},
  \citenamefont {Harry},\ and\ \citenamefont {Ogin}}]{eva08}%
  \BibitemOpen
  \bibfield  {author} {\bibinfo {author} {\bibfnamefont {M.}~\bibnamefont
  {Evans}}, \bibinfo {author} {\bibfnamefont {S.}~\bibnamefont {Ballmer}},
  \bibinfo {author} {\bibfnamefont {M.}~\bibnamefont {Fejer}}, \bibinfo
  {author} {\bibfnamefont {P.}~\bibnamefont {Fritschel}}, \bibinfo {author}
  {\bibfnamefont {G.}~\bibnamefont {Harry}},\ and\ \bibinfo {author}
  {\bibfnamefont {G.}~\bibnamefont {Ogin}},\ }\bibfield  {title} {\bibinfo
  {title} {Thermo-optic noise in coated mirrors for high-precision optical
  measurements},\ }\href {https://doi.org/10.1103/PhysRevD.78.102003}
  {\bibfield  {journal} {\bibinfo  {journal} {Phys. Rev. D}\ }\textbf {\bibinfo
  {volume} {78}},\ \bibinfo {pages} {102003} (\bibinfo {year}
  {2008})}\BibitemShut {NoStop}%
\bibitem [{\citenamefont {Chalermsongsak}\ \emph {et~al.}(2016)\citenamefont
  {Chalermsongsak}, \citenamefont {Hall}, \citenamefont {Cole}, \citenamefont
  {Follman}, \citenamefont {Seifert}, \citenamefont {Arai}, \citenamefont
  {Gustafson}, \citenamefont {Smith}, \citenamefont {Aspelmeyer},\ and\
  \citenamefont {Adhikari}}]{cha16}%
  \BibitemOpen
  \bibfield  {author} {\bibinfo {author} {\bibfnamefont {T.}~\bibnamefont
  {Chalermsongsak}}, \bibinfo {author} {\bibfnamefont {E.~D.}\ \bibnamefont
  {Hall}}, \bibinfo {author} {\bibfnamefont {G.~D.}\ \bibnamefont {Cole}},
  \bibinfo {author} {\bibfnamefont {D.}~\bibnamefont {Follman}}, \bibinfo
  {author} {\bibfnamefont {F.}~\bibnamefont {Seifert}}, \bibinfo {author}
  {\bibfnamefont {K.}~\bibnamefont {Arai}}, \bibinfo {author} {\bibfnamefont
  {E.~K.}\ \bibnamefont {Gustafson}}, \bibinfo {author} {\bibfnamefont {J.~R.}\
  \bibnamefont {Smith}}, \bibinfo {author} {\bibfnamefont {M.}~\bibnamefont
  {Aspelmeyer}},\ and\ \bibinfo {author} {\bibfnamefont {R.~X.}\ \bibnamefont
  {Adhikari}},\ }\bibfield  {title} {\bibinfo {title} {Coherent cancellation of
  photothermal noise in {GaAs/Al$_{0.92}$Ga$_{0.08}$As} {B}ragg mirrors},\
  }\href {https://doi.org/10.1088/0026-1394/53/2/860} {\bibfield  {journal}
  {\bibinfo  {journal} {Metrologia}\ }\textbf {\bibinfo {volume} {53}},\
  \bibinfo {pages} {860} (\bibinfo {year} {2016})}\BibitemShut {NoStop}%
\bibitem [{\citenamefont {Cerdonio}\ \emph {et~al.}(2001)\citenamefont
  {Cerdonio}, \citenamefont {Conti}, \citenamefont {Heidmann},\ and\
  \citenamefont {Pinard}}]{cer01}%
  \BibitemOpen
  \bibfield  {author} {\bibinfo {author} {\bibfnamefont {M.}~\bibnamefont
  {Cerdonio}}, \bibinfo {author} {\bibfnamefont {L.}~\bibnamefont {Conti}},
  \bibinfo {author} {\bibfnamefont {A.}~\bibnamefont {Heidmann}},\ and\
  \bibinfo {author} {\bibfnamefont {M.}~\bibnamefont {Pinard}},\ }\bibfield
  {title} {\bibinfo {title} {Thermoelastic effects at low temperatures and
  quantum limits in displacement measurements},\ }\href
  {https://doi.org/10.1103/PhysRevD.63.082003} {\bibfield  {journal} {\bibinfo
  {journal} {Phys. Rev. D}\ }\textbf {\bibinfo {volume} {63}},\ \bibinfo
  {pages} {082003} (\bibinfo {year} {2001})}\BibitemShut {NoStop}%
\bibitem [{\citenamefont {Zhang}\ \emph {et~al.}(2014)\citenamefont {Zhang},
  \citenamefont {Martin}, \citenamefont {Benko}, \citenamefont {Hall},
  \citenamefont {Ye}, \citenamefont {Hagemann}, \citenamefont {Legero},
  \citenamefont {Sterr}, \citenamefont {Riehle}, \citenamefont {Cole},\ and\
  \citenamefont {Aspelmeyer}}]{zha14}%
  \BibitemOpen
  \bibfield  {author} {\bibinfo {author} {\bibfnamefont {W.}~\bibnamefont
  {Zhang}}, \bibinfo {author} {\bibfnamefont {M.~J.}\ \bibnamefont {Martin}},
  \bibinfo {author} {\bibfnamefont {C.}~\bibnamefont {Benko}}, \bibinfo
  {author} {\bibfnamefont {J.~L.}\ \bibnamefont {Hall}}, \bibinfo {author}
  {\bibfnamefont {J.}~\bibnamefont {Ye}}, \bibinfo {author} {\bibfnamefont
  {C.}~\bibnamefont {Hagemann}}, \bibinfo {author} {\bibfnamefont
  {T.}~\bibnamefont {Legero}}, \bibinfo {author} {\bibfnamefont
  {U.}~\bibnamefont {Sterr}}, \bibinfo {author} {\bibfnamefont
  {F.}~\bibnamefont {Riehle}}, \bibinfo {author} {\bibfnamefont {G.~D.}\
  \bibnamefont {Cole}},\ and\ \bibinfo {author} {\bibfnamefont
  {M.}~\bibnamefont {Aspelmeyer}},\ }\bibfield  {title} {\bibinfo {title}
  {Reduction of residual amplitude modulation to $1\times10^{-6}$ for
  frequency-modulation and laser stabilization},\ }\href
  {https://doi.org/10.1364/OL.39.001980} {\bibfield  {journal} {\bibinfo
  {journal} {Opt. Lett.}\ }\textbf {\bibinfo {volume} {39}},\ \bibinfo {pages}
  {1980} (\bibinfo {year} {2014})}\BibitemShut {NoStop}%
\bibitem [{\citenamefont {Yu}(2023)}]{yu23}%
  \BibitemOpen
  \bibfield  {author} {\bibinfo {author} {\bibfnamefont {J.}~\bibnamefont
  {Yu}},\ }\href {https://doi.org/10.15488/13416} {\bibinfo {title} {Cryogenic
  silicon {Fabry}-{Perot} resonator with \\ {Al$_{0.92}$Ga$_{0.08}$As/GaAs}
  mirror coatings.}},\ \bibinfo {howpublished} {Ph.D. thesis,
  QUEST-Leibniz-Forschungsschule der Gottfried Wilhelm Leibniz Universit\"at
  Hannover} (\bibinfo {year} {2023})\BibitemShut {NoStop}%
\bibitem [{\citenamefont {Barbarat}\ \emph {et~al.}(2025)\citenamefont
  {Barbarat}, \citenamefont {Benkler}, \citenamefont {Bober}, \citenamefont
  {Clivati}, \citenamefont {Dickmann}, \citenamefont {Fang}, \citenamefont
  {Fluhr}, \citenamefont {Fordell}, \citenamefont {Gillot}, \citenamefont
  {Giordano}, \citenamefont {Gustavsson}, \citenamefont {Hanhij\"arvi},
  \citenamefont {Hartman}, \citenamefont {Herbers}, \citenamefont {Jaros},
  \citenamefont {Kawohl}, \citenamefont {Kersal\'e}, \citenamefont {Kroker},
  \citenamefont {Kwong}, \citenamefont {Lacro\^ute}, \citenamefont {Targat},
  \citenamefont {Legero}, \citenamefont {Lind\'en}, \citenamefont {Lindvall},
  \citenamefont {Lodewyck}, \citenamefont {Ma}, \citenamefont {Meyer},
  \citenamefont {Morzy\'nski}, \citenamefont {Millo}, \citenamefont
  {Naro\.znik}, \citenamefont {Nicolodi}, \citenamefont {Penza}, \citenamefont
  {Pointard}, \citenamefont {Rippe}, \citenamefont {Risaro}, \citenamefont
  {Roset}, \citenamefont {Sauer}, \citenamefont {Savio}, \citenamefont
  {Schiller}, \citenamefont {Neto}, \citenamefont {Sterr}, \citenamefont
  {Vartehparvar}, \citenamefont {Vogt}, \citenamefont {Wagner}, \citenamefont
  {Wallin}, \citenamefont {Wiens}, \citenamefont {Yu}, \citenamefont {Zawada},\
  and\ \citenamefont {Zelan}}]{bar25}%
  \BibitemOpen
  \bibfield  {author} {\bibinfo {author} {\bibfnamefont {J.}~\bibnamefont
  {Barbarat}}, \bibinfo {author} {\bibfnamefont {E.}~\bibnamefont {Benkler}},
  \bibinfo {author} {\bibfnamefont {M.}~\bibnamefont {Bober}}, \bibinfo
  {author} {\bibfnamefont {C.}~\bibnamefont {Clivati}}, \bibinfo {author}
  {\bibfnamefont {J.}~\bibnamefont {Dickmann}}, \bibinfo {author}
  {\bibfnamefont {B.}~\bibnamefont {Fang}}, \bibinfo {author} {\bibfnamefont
  {C.}~\bibnamefont {Fluhr}}, \bibinfo {author} {\bibfnamefont
  {T.}~\bibnamefont {Fordell}}, \bibinfo {author} {\bibfnamefont
  {J.}~\bibnamefont {Gillot}}, \bibinfo {author} {\bibfnamefont
  {V.}~\bibnamefont {Giordano}}, \bibinfo {author} {\bibfnamefont
  {D.}~\bibnamefont {Gustavsson}}, \bibinfo {author} {\bibfnamefont
  {K.}~\bibnamefont {Hanhij\"arvi}}, \bibinfo {author} {\bibfnamefont
  {M.}~\bibnamefont {Hartman}}, \bibinfo {author} {\bibfnamefont
  {S.}~\bibnamefont {Herbers}}, \bibinfo {author} {\bibfnamefont
  {A.}~\bibnamefont {Jaros}}, \bibinfo {author} {\bibfnamefont
  {J.}~\bibnamefont {Kawohl}}, \bibinfo {author} {\bibfnamefont
  {Y.}~\bibnamefont {Kersal\'e}}, \bibinfo {author} {\bibfnamefont
  {S.}~\bibnamefont {Kroker}}, \bibinfo {author} {\bibfnamefont {C.~J.}\
  \bibnamefont {Kwong}}, \bibinfo {author} {\bibfnamefont {C.}~\bibnamefont
  {Lacro\^ute}}, \bibinfo {author} {\bibfnamefont {R.~L.}\ \bibnamefont
  {Targat}}, \bibinfo {author} {\bibfnamefont {T.}~\bibnamefont {Legero}},
  \bibinfo {author} {\bibfnamefont {M.}~\bibnamefont {Lind\'en}}, \bibinfo
  {author} {\bibfnamefont {T.}~\bibnamefont {Lindvall}}, \bibinfo {author}
  {\bibfnamefont {J.}~\bibnamefont {Lodewyck}}, \bibinfo {author}
  {\bibfnamefont {C.~Y.}\ \bibnamefont {Ma}}, \bibinfo {author} {\bibfnamefont
  {R.}~\bibnamefont {Meyer}}, \bibinfo {author} {\bibfnamefont
  {P.}~\bibnamefont {Morzy\'nski}}, \bibinfo {author} {\bibfnamefont
  {J.}~\bibnamefont {Millo}}, \bibinfo {author} {\bibfnamefont
  {M.}~\bibnamefont {Naro\.znik}}, \bibinfo {author} {\bibfnamefont
  {D.}~\bibnamefont {Nicolodi}}, \bibinfo {author} {\bibfnamefont
  {A.}~\bibnamefont {Penza}}, \bibinfo {author} {\bibfnamefont
  {B.}~\bibnamefont {Pointard}}, \bibinfo {author} {\bibfnamefont
  {L.}~\bibnamefont {Rippe}}, \bibinfo {author} {\bibfnamefont
  {M.}~\bibnamefont {Risaro}}, \bibinfo {author} {\bibfnamefont
  {P.}~\bibnamefont {Roset}}, \bibinfo {author} {\bibfnamefont
  {S.}~\bibnamefont {Sauer}}, \bibinfo {author} {\bibfnamefont
  {P.}~\bibnamefont {Savio}}, \bibinfo {author} {\bibfnamefont
  {S.}~\bibnamefont {Schiller}}, \bibinfo {author} {\bibfnamefont {L.~S.}\
  \bibnamefont {Neto}}, \bibinfo {author} {\bibfnamefont {U.}~\bibnamefont
  {Sterr}}, \bibinfo {author} {\bibfnamefont {O.}~\bibnamefont {Vartehparvar}},
  \bibinfo {author} {\bibfnamefont {V.}~\bibnamefont {Vogt}}, \bibinfo {author}
  {\bibfnamefont {N.}~\bibnamefont {Wagner}}, \bibinfo {author} {\bibfnamefont
  {A.~E.}\ \bibnamefont {Wallin}}, \bibinfo {author} {\bibfnamefont
  {E.}~\bibnamefont {Wiens}}, \bibinfo {author} {\bibfnamefont
  {J.}~\bibnamefont {Yu}}, \bibinfo {author} {\bibfnamefont {M.}~\bibnamefont
  {Zawada}},\ and\ \bibinfo {author} {\bibfnamefont {M.}~\bibnamefont
  {Zelan}},\ }\href {https://doi.org/10.48550/arXiv.2504.06213} {\bibinfo
  {title} {Guidelines for designs for ultrastable laser with $10^{-17}$
  fractional frequency instability}},\ \bibinfo {howpublished}
  {arXiv:2504.06213 [physics.optics]} (\bibinfo {year} {2025})\BibitemShut
  {NoStop}%
\bibitem [{\citenamefont {Freier}\ \emph {et~al.}(2016)\citenamefont {Freier},
  \citenamefont {Hauth}, \citenamefont {Schkolnik}, \citenamefont {Leykauf},
  \citenamefont {Schilling}, \citenamefont {Wziontek}, \citenamefont
  {Scherneck}, \citenamefont {M\"uller},\ and\ \citenamefont {Peters}}]{fre16}%
  \BibitemOpen
  \bibfield  {author} {\bibinfo {author} {\bibfnamefont {C.}~\bibnamefont
  {Freier}}, \bibinfo {author} {\bibfnamefont {M.}~\bibnamefont {Hauth}},
  \bibinfo {author} {\bibfnamefont {V.}~\bibnamefont {Schkolnik}}, \bibinfo
  {author} {\bibfnamefont {B.}~\bibnamefont {Leykauf}}, \bibinfo {author}
  {\bibfnamefont {M.}~\bibnamefont {Schilling}}, \bibinfo {author}
  {\bibfnamefont {H.}~\bibnamefont {Wziontek}}, \bibinfo {author}
  {\bibfnamefont {H.-G.}\ \bibnamefont {Scherneck}}, \bibinfo {author}
  {\bibfnamefont {J.}~\bibnamefont {M\"uller}},\ and\ \bibinfo {author}
  {\bibfnamefont {A.}~\bibnamefont {Peters}},\ }\bibfield  {title} {\bibinfo
  {title} {Mobile quantum gravity sensor with unprecedented stability},\ }\href
  {https://doi.org/10.1088/1742-6596/723/1/012050} {\bibfield  {journal}
  {\bibinfo  {journal} {J. Phys.: Conf. Ser.}\ }\textbf {\bibinfo {volume}
  {723}},\ \bibinfo {pages} {012050} (\bibinfo {year} {2016})}\BibitemShut
  {NoStop}%
\bibitem [{\citenamefont {Perner}\ \emph {et~al.}(2024)\citenamefont {Perner},
  \citenamefont {Winkler}, \citenamefont {Truong}, \citenamefont {Zhao},
  \citenamefont {Bachmann}, \citenamefont {Mayer}, \citenamefont {Fellinger},
  \citenamefont {Follman}, \citenamefont {Heu}, \citenamefont {Deutsch},
  \citenamefont {Bailey}, \citenamefont {Peelaers}, \citenamefont {Puchegger},
  \citenamefont {Fleisher}, \citenamefont {Cole},\ and\ \citenamefont
  {Heckl}}]{per24}%
  \BibitemOpen
  \bibfield  {author} {\bibinfo {author} {\bibfnamefont {L.~W.}\ \bibnamefont
  {Perner}}, \bibinfo {author} {\bibfnamefont {G.}~\bibnamefont {Winkler}},
  \bibinfo {author} {\bibfnamefont {G.-W.}\ \bibnamefont {Truong}}, \bibinfo
  {author} {\bibfnamefont {G.}~\bibnamefont {Zhao}}, \bibinfo {author}
  {\bibfnamefont {D.}~\bibnamefont {Bachmann}}, \bibinfo {author}
  {\bibfnamefont {A.~S.}\ \bibnamefont {Mayer}}, \bibinfo {author}
  {\bibfnamefont {J.}~\bibnamefont {Fellinger}}, \bibinfo {author}
  {\bibfnamefont {D.}~\bibnamefont {Follman}}, \bibinfo {author} {\bibfnamefont
  {P.}~\bibnamefont {Heu}}, \bibinfo {author} {\bibfnamefont {C.}~\bibnamefont
  {Deutsch}}, \bibinfo {author} {\bibfnamefont {D.~M.}\ \bibnamefont {Bailey}},
  \bibinfo {author} {\bibfnamefont {H.}~\bibnamefont {Peelaers}}, \bibinfo
  {author} {\bibfnamefont {S.}~\bibnamefont {Puchegger}}, \bibinfo {author}
  {\bibfnamefont {A.~J.}\ \bibnamefont {Fleisher}}, \bibinfo {author}
  {\bibfnamefont {G.~D.}\ \bibnamefont {Cole}},\ and\ \bibinfo {author}
  {\bibfnamefont {O.~H.}\ \bibnamefont {Heckl}},\ }\bibfield  {title} {\bibinfo
  {title} {Mid-infrared interference coatings with excess optical loss below
  10~ppm: erratum},\ }\href {https://doi.org/10.1364/OPTICA.520398} {\bibfield
  {journal} {\bibinfo  {journal} {Optica}\ }\textbf {\bibinfo {volume} {11}},\
  \bibinfo {pages} {619} (\bibinfo {year} {2024})}\BibitemShut {NoStop}%
\end{thebibliography}%
\clearpage
\appendix

\end{document}